\documentclass[letterpaper,twocolumn,10pt]{article}
\usepackage{usenix-2020-09}

\usepackage{tikz}
\usepackage{amsmath}
\usepackage{amssymb}
\usepackage{pifont}
\usepackage{filecontents}

\usepackage{graphicx}
\usepackage{multirow}
\usepackage{booktabs}
\usepackage{amssymb}
\usepackage{subfig}
\usepackage{enumitem}
\usepackage{array} 
\usepackage{xspace}
\usepackage{comment}

\usepackage{algpseudocode}
\usepackage{algorithm}
\usepackage{authblk}

\newcommand\sysname{\textsf{Varuna}\xspace}

\usepackage{balance}
\usepackage{titling}
\usepackage{setspace}
\pretitle{\vspace*{-0.8cm}\begin{center}\bfseries\Large}
\posttitle{\end{center}\vspace{-0.8em}}
\preauthor{\begin{center}\normalsize}
\postauthor{\end{center}\vspace{-1.0em}}
\predate{}\postdate{}

\begin{document}

\date{}

\title{\Large \bf Varuna: Enabling Failure-Type Aware RDMA Failover}
\author{
{\rm Xiaoyang Wang$^1$, Yongkun Li$^{1*}$, Lulu Yao$^2$, Guoli Wei$^1$, Longcheng Yang$^1$, Yinlong Xu$^1$ \\ \vspace{-7pt}Weiqing Kong$^3$, Weiguang Wang$^3$, Peng Dong$^3$, Bingyang Liu$^3$}
\vspace{-3pt}
\\
\textit{$^1$University of Science and Technology of China}\quad
\textit{$^2$Ningbo University}\quad
\textit{$^3$Huawei}
}


\maketitle

\begin{abstract}





RDMA link failures can render connections temporarily unavailable, causing both performance degradation and significant recovery overhead.
To tolerate such failures, production datacenters assign each primary link with a standby link and, upon failure, uniformly retransmit all in-flight RDMA request over the backup path.
However, we observe that such blanket retransmission is unnecessary. In-flight requests can be split into pre-failure and post-failure categories depending on whether the responder has already executed.
Retransmitting post-failure requests is not only redundant—consuming bandwidth—but also incorrect for non-idempotent operations, where duplicate execution can violate application semantics.

We present \sysname, a failure-type-aware RDMA recovery mechanism that enables correct retransmission and µs-level failover.
\sysname piggybacks a lightweight completion log on every RDMA operation; after a link failure, this log deterministically reveals which in-flight requests were executed (post-failure) and which were lost (pre-failure).
\sysname then retransmits only the pre-failure subset and fetches/recovers the return values for post-failure requests.
Evaluated using synthetic microbenchmarks and end-to-end RDMA TPC-C transactions, \sysname incurs only 0.6–10\% steady-state latency overhead in realistic applications, eliminates 65\% of recovery retransmission time, preserves transactional consistency, and introduces zero connectivity rebuild overhead and negligible memory overhead during RDMA failover.



\end{abstract}

\footnotetext[1]{Yongkun Li is the corresponding author.}

\section{Introduction}
\label{sec:intro}






The growing prevalence of big data and large-scale AI models has driven a critical need for high-throughput, low-latency networks in data centers and distributed clusters to enable efficient data communication~\cite{storageRDMA22,lin2024luberdma}. RDMA (\underline{R}emote \underline{D}irect \underline{M}emory \underline{A}ccess)~\cite{rdma}, with its high bandwidth, ultra-low latency, and minimal CPU overhead, has thus become essential infrastructure for building modern distributed systems such as key-value stores~\cite{fusee,motor} and AI applications~\cite{megascale, hpn7, mooncake}. By employing mechanisms including CPU bypass, kernel bypass, zero-copy operations, and direct remote memory access, the latest NICs can deliver bandwidths of up to 800 Gbps and nanosecond-scale latency~\cite{mlx8}.

Despite these advantages, RDMA deployments increasingly suffer from single-point RDMA failures—including link failures and link flappings—as application and cluster scales continue to grow. Link flapping, in particular, has become common: when a link flaps, it repeatedly goes down and typically recovers only after several seconds\cite{hpn7, bytetracker,megascale}. Operational reports highlight the severity of the problem. For example, Alibaba observes 5,000-60,000 link flaps per day in a cluster with 15,000 GPUs, while the monthly link failure rate reaches 0.057\%. 
In another 3,000-GPU cluster, a single link failure can incur losses of up to 30,000 dollars~\cite{hpn7}. 

As a result, RDMA failures increasingly degrade application performance at scale~\cite{megascale, hpn7, mprdma, hostping, bytetracker,mooncake,luberdma,hostmesh}. For transactional services such as key-value stores, RDMA failures can significantly reduce throughput when certain nodes become temporarily unavailable, and may even lead to data inconsistency~\cite{motor, ukharon}. For highly parallel tasks like AI training, a single RDMA failure may cause an entire job to abort. To recover from RDMA network errors, applications  typically rely on software-level strategies such as checkpointing, application migration, or log replay. However, these approaches are costly—often taking from seconds to hours—and they fundamentally cannot address the temporary unavailability of RDMA-attached resources, leading to extended downtime and resource underutilization\cite{luberdma, ByteCheckpoint25, hpn7}.

To address RDMA failures, datacenters deploy hardware-level redundancy—such as multiple network ports\cite{hpn7}, NICs\cite{oasis, luberdma, mooncake}, and switches\cite{hpn7}—to provide alternative transmission links. These backup RDMA paths can handle both persistent link failures and transient link flapping, thereby reducing the overall effect of RDMA failure. In practice, whether the failure is a hard link outage or a brief flap, the common recovery strategy is to switch traffic to a standby link and retransmit the in-flight RDMA requests.

However, link-switchover–based retransmission is far from a silver bullet. A closer examination of RDMA link failures reveals that not all failed requests actually require retransmission. We find that in-flight requests naturally fall into two categories depending on whether the responder has already executed them: \textit{pre-failure} requests, which were lost before reaching the responder and must be retransmitted, and \textit{post-failure} requests, whose responses were lost after the responder completed the operation.

Blindly retransmitting post-failure requests is problematic for two reasons. First, it wastes external bandwidth by redundantly resending data that has already been applied at the remote end. More critically, it jeopardizes data correctness for non-idempotent operations. An in-flight non-idempotent verb (e.g., Write, Compare-and-Swap, Fetch-and-Add) may have completed on the responder even though its acknowledgment was lost. Retransmitting such an operation to the backup NIC re-executes it and corrupts remote state. Our experiments show that under high-bandwidth conditions, blanket retransmission can increase end-to-end transmission time by approximately 182.1\%, and retransmissions of post-failure requests can cause up to 83.9\% of transactional inconsistencies.

Furthermore, multi-link failover suffers from a fundamental performance–space trade-off. Re-establishing communication contexts (e.g., Reliable Connection Queue Pairs or RCQPs) for the backup link incurs substantial computation and multi-second stalls, whereas pre-allocating these contexts for fast failover doubles QP memory consumption in practice. Both options impose significant overheads on modern deployments.

We propose \sysname, a failure-type–aware RDMA recovery mechanism that ensures correct retransmission and low-cost, fast failover.
    \sysname piggybacks a lightweight completion log on every RDMA operation; upon a link failure, this log deterministically identifies which in-flight requests were executed (post-failure) and which were lost (pre-failure). 
    \sysname retransmits only the pre-failure subset and retrieves the return values of post-failure requests to avoid redundant or incorrect re-execution.
    To minimize failover latency, \sysname instantiates Dynamic Connection Queue Pairs (DCQPs) on standby links and introduces a high-performance, application-transparent RCQP–DCQP co-working link-switch framework that integrates lightweight failure coherency and recovery. 

Specifically, the design of \sysname is as follows: 

\textbf{1) Log-based Fail Recognition.}
To distinguish between different types of failures, \sysname maintains logs on both the requester and the responser sides.
In particular, for each non‑idempotent operation, the requester writes to its local log and simultaneously appends a compact 
completion record using an inline one‑sided RDMA write to a small log entry on the remote peer.
The completion state is recorded independently of RDMA request acknowledgments, thus \sysname can consult these logs after a failover to 
determine exactly which requests executed and which must be retransmitted.

\textbf{2) Extended Status in Post Failure.}
To guarantee the data correctness of post failure, \sysname carries a unique per‑operation identifier (UID) and the operation’s result is recorded alongside that UID for each request. \sysname uses UIDs to deterministically detect whether a given atomic operation completed during recovery 
and to recover its outcome without ambiguity.

\textbf{3) Lightweight Fast recovery.}
\sysname maintains a small, configurable pool of lightweight DCQPs on each NIC that can be shared across endpoints on the same host to different targets. When a link fails, traffic is redirected to DCQPs on a standby link so communication can resume within milliseconds. All RCQPs are rebuilt asynchronously and swapped in later, avoiding the memory blowup associated with pre‑caching RCQPs on every link to enable lightweight fast recovery.

We evaluate \sysname on multi‑NIC servers using representative microbenchmarks and end‑to‑end RDMA transactions with TPC-C workloads. \sysname incurs 0.6\%-10\% steady-state latency overhead, eliminates 65\% of retransmission traffic, preserves consistency transactional with zero connectivity rebuild overhead and negligible memory overhead.





The rest of the paper is organized as follows. We first examine the limitations of existing RDMA failover mechanisms (Sec.\ref{sec:motivation}), then describe the design of \sysname and implementation (Secs.\ref{sec:design},\ref{sec:impl}), and evaluate its recovery performance using benchmarks and end-to-end applications (Sec.\ref{sec:evaluation}). We discuss limitations and potential extensions (Sec.\ref{sec:limitation}), review related work (Sec.\ref{sec:relatedwork}), and conclude (Sec.~\ref{sec:conclusion}). The implementation of \sysname is available at GitHub~\footnote{https://github.com/Plasticinew/Varuna}.

\section{Background and Motivation}
\label{sec:motivation}

\subsection{RDMA with Failure}
\noindent\textbf{\underline{Effectiveness of RDMA.}} 
RDMA has become the de facto communication primitive for high‑performance datacenter services. By exposing RDMA verbs and Queue Pairs (QPs), it supports one‑sided operations (READ, WRITE, and atomic operations) as well as two‑sided send/receive semantics that bypass the kernel and avoid extra data copies. The result is microsecond‑level latencies, low CPU utilization, and near line‑rate throughput—properties that make RDMA central to systems such as distributed key‑value stores\cite{fusee}, remote memory/disaggregation platforms~\cite{legoos, fastswap, infiniswap}, and large‑scale AI training\cite{hpn7} and serving\cite{mooncake}.

\noindent\textbf{\underline{RDMA failure.}} 
As application scales continuous grow, an increasing number of hosts are connected with RDMA to execute distributed tasks, which in turn raises the risk of single‑point RDMA failures\cite{megascale, hpn7, mprdma, hostping, bytetracker,mooncake,luberdma,hostmesh}.
The single-point RDMA failures primarily include the following three categories: link failure, link flapping, and NIC failure.
(i) Link failure refers to the physical link of an RDMA network card remaining persistently and stably in the "DOWN" state, unable to establish a physical connection.
(ii) Link flapping refers to the frequent, intermittent switching of the physical link (e.g., Ethernet or InfiniBand port) of an RDMA between the "UP" (connected) and "DOWN" (disconnected) states, such as several times per second or per minute.
(iii) NIC failure refers to the RDMA network card itself exhibiting functional abnormalities or a complete failure, which extends beyond the scope of a single port.

\noindent\textbf{\underline{Impact of RDMA failure.}} 
In pursuit of extreme performance, RDMA sacrifices the resilience and buffering layers of the traditional network stack, shifting the complexity of network reliability from software to hardware and physical infrastructure. Consequently, the cost and impact of RDMA failures far exceed those of traditional TCP/IP networks.
For transactional applications (such as key-value stores), RDMA failures typically severely degrade their performance, lead to service unavailability, and may even compromise data consistency. Like Motor and uKharon~\cite{motor, ukharon}, when network failures occur, they assume that RDMA failures are detected and resolved by data center administrators. According to the CAP theorem~\cite{CAPtheory1,CAPtheory2}, either availability or consistency cannot be fully guaranteed. 
For highly parallel tasks, such as high performance computing and AI training/inference, RDMA failures can range from causing performance degradation to rendering training unavailable. For example, in Alibaba clusters, 5,000 to 60,000 occurrences of link flapping happen daily, resulting in temporary performance degradation~\cite{hpn7}.

\subsection{Recovery Techniques of RDMA failure}
\noindent\textbf{\underline{Recovery with Application Aware.}}
Current distributed applications typically employ \textit{fault recovery} and \textit{fault tolerance} to address network failures. \textit{Fault recovery} mechanisms for applications often utilize techniques such as checkpointing~\cite{GeminiRecovery23,CheckFreq23,CheckNRun23,FlowCheck25,ByteCheckpoint25} or logging~\cite{Lineagestash19,luo2024splitft,qi2025efficient} to restore failed applications and minimize overhead. However, these approaches not only introduce additional storage overhead but also incur computational costs and latency due to log replay or checkpoint recovery, thereby increasing costs~\cite{xu2025resource,thorpe2023bamboo}.

\textit{Fault tolerance} in applications primarily aims to prevent the failure of a single worker—often caused by network issues—from rendering the entire application unavailable. It's typically achieved by replacing the faulty node with a new one~\cite{jang2023oobleck,duan2024efficient,zhuang2021hoplite}. For instance, in distributed key-value storage systems, when the primary node responsible for write operations becomes unavailable, the system must undergo a multi-step process: leader election, link restoration, log replay, and resumption of new write operations. This significantly degrades both the availability of the service and the performance of the application.

While implementing fault recovery and fault tolerance mechanisms at the application layer mitigate the availability impact caused by RDMA failures, it inevitably incurs significant performance overhead, making it difficult for the system to meet predefined service level objects (SLOs) requirements.

\noindent\textbf{\underline{Recovery with RDMA Link-switchover.}}
To better address RDMA failures, modern data centers widely adopt a multi-path redundant, high-availability network architecture. By deploying multiple physical links between servers and configuring them to serve as mutual backups, rapid link-switchover can be achieved in the event of link or switch failures~\cite{luberdma}. The link-switchover strategy typically relies on servers equipped with multiple network ports, multiple NICs, or multiple switches. Currently, such RDMA recovery design based on link-layer redundancy and fast rerouting has become an infrastructure standard for data center networks~\cite{bai2023empowering,hpn7,hu2024characterization,gill2011understanding,guo2016rdma}.

In practice, regardless of whether a failure is a hard link drop or a brief disruption, the prevailing recovery strategy is to shift traffic onto a standby path (or backup path) and retransmit all in-flight RDMA requests. Throughout this paper, we use backup link as an umbrella term encompassing a variety of redundancy architectures, including multi-port bonding, multi-NIC configurations (within a single node or pooled across nodes via CXL like Oasis\cite{oasis} discussed), and multi-plane RDMA switch fabrics.

The link-switch and retransmission–based RDMA failover framework—exemplified by LubeRDMA \cite{luberdma} and Mooncake Transfer Engine \cite{mooncake}—operates at the RDMA network–library layer, between the application and the NIC hardware. This placement allows the framework to maintain a global view of multiple links and to support a wide range of applications. Upon a link failure, the framework consults its local list of in-flight RDMA requests and uniformly retransmits all of them to the responder.

\subsection{Link-switchover-based Retransmission}
This subsection aims to conduct an in-depth analysis of retransmission mechanisms based on link switchover. Existing strategies typically adopt an indiscriminate and blind retransmission pattern, lacking precise awareness of the specific transmission stage in which a failure occurs. To address this, we systematically analyze the actual transmission stages where retransmission operations take effect and, based on this, redefine a more targeted design space for retransmission.

\begin{figure}[!t]
 \centering
 \includegraphics[width=0.34\textwidth]{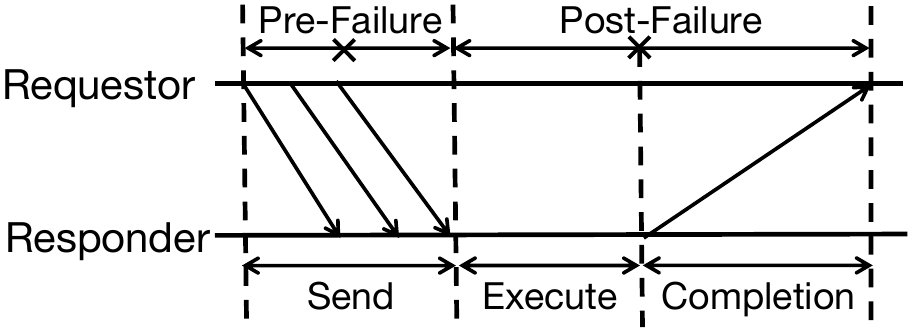}
 \caption{Transmission Stages and Failure Types: 
 If a network failure occurs before execution, the entire request is considered un-executed (pre-failure). Otherwise, the request execution is completed at the responder, but the ACK may be lost (post-failure), and no retransmission is required.}
 \vskip -10pt
 \label{fig:moti_progress}
\vskip -6pt
\end{figure}

As shown in Fig.~\ref{fig:moti_progress}, an RDMA request proceeds through a simple pipeline according to RDMA specifications. (i) The requester generates packets according to its work request and transmits them to the responder. (ii) After receiving and verifying the complete packet sequence, the responder executes the RDMA operation. (iii) The responder returns an acknowledgment (ACK) once the operation finishes, signaling completion to the requester.

A critical point in this pipeline is the execution commit at the responder. Execution is assumed atomic—once started, it cannot be partially applied. Therefore, from the requestor’s perspective, an RDMA request can be divided into two stages: i) Send Stage: packets are in flight and the responder has not yet executed the request. ii) Completion Stage: the responder has executed the request and is sending the ACK back.
If an RDMA failure occurs during the Send Stage, the request has not been executed; we classify this as \textit{pre-failure}.
If the failure occurs after the responder has fully received and executed the request—whether before or during the ACK transmission—the requester will still observe the request as incomplete, but the operation has already taken effect at the responder. We classify this scenario as \textit{post-failure}.

In summary, from the requestor’s viewpoint, both cases look identical—no ACK returned. However, the responder’s execution state distinguishes non-executed (pre-failure) from already executed (post-failure). Misclassifying post-failure requests leads to duplicate execution on retransmission.

\begin{figure}[!t]
 \centering
 \includegraphics[width=0.4\textwidth]{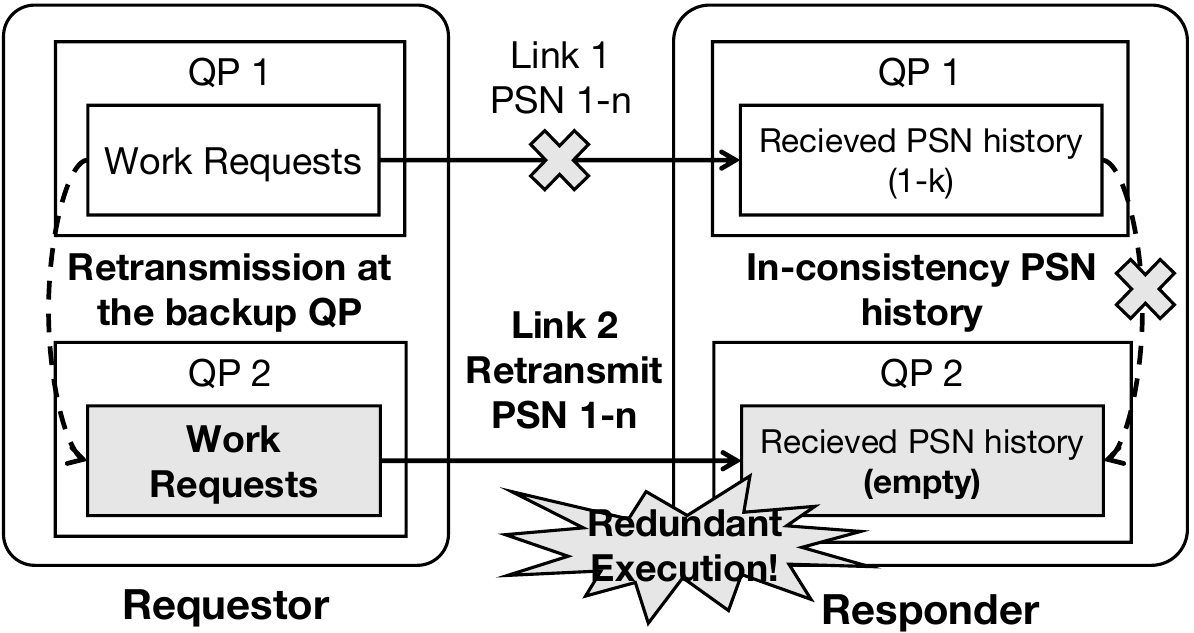}
 \caption{Retransmission Flow: When retransmitting using a backup link with a backup QP, the standard RDMA retransmission detection mechanism is bypassed. As a result, previously transmitted requests may be applied again at the receiver, leading to redundant execution.}
\vskip -10pt
 \label{fig:retransmission}
\end{figure}

\begin{figure}[!t]
 \centering
   \subfloat[Post-Failure Ratio.\label{fig:postfailure}]{
    \includegraphics[scale=0.4]{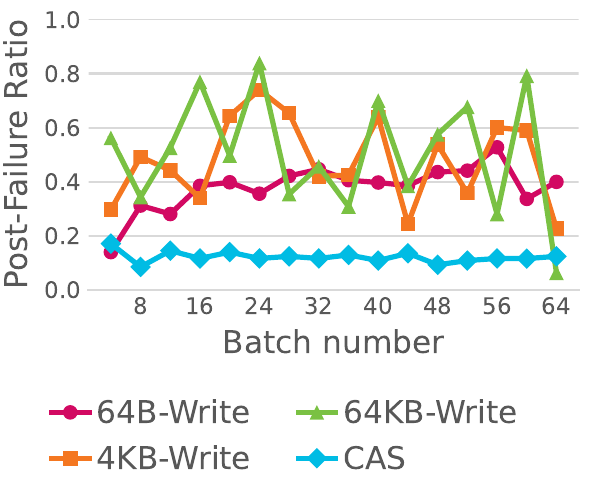}
 }
 \subfloat[Uniform Resend Overhead. \label{fig:motivate_resend}]{
    \includegraphics[scale=0.4]{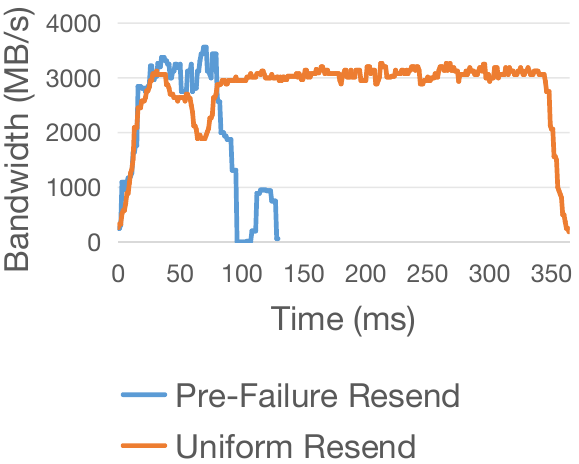}
 }
  \vskip -7pt
 \caption{Post-Failure in RDMA:
(a) Post-failure occurrences are unpredictable across different workloads.
(b) Identifying post-failure requests reduces unnecessary resend overhead.}\label{fig:motivation} 
 \vskip -5pt
\end{figure}

\subsection{Limitations with Retransmission}
In general, RDMA’s history packet sequence numbers (PSNs) record and ACK-buffer design can suppress duplicate execution within the same QP by re-emitting cached ACKs for repeated PSNs. However, if the recovery mechanism focuses on a more general scenario where the original QP may be unusable after a failure and the backup link may reside on another port or even another NIC with a fresh new QP. In such cases, the built-in RDMA retransmission mechanism is insufficient, as illustrated in Fig.~\ref{fig:retransmission}. 
In particular, blind retransmission strategies has several limitations: 

\textbf{1) Significant redundant retransmission overhead.} Blind retransmission resend every in-flight request, even though only the pre-failure subset actually requires retransmission. This leads to unnecessary bandwidth consumption and extended recovery latency. To quantify the necessity of distinguishing pre- and post-failure requests, we evaluate RDMA one-sided operations using 16 client threads against a single server, covering a range of payload sizes—from an 8 B Compare-and-Swap (CAS), to a 64 B small write, to a 4 KB write, to a 64 KB large write. We vary the send-batch size (i.e., the number of packets aggregated before generating one completion) and inject failures by manually bringing the RDMA NIC port down.

As shown in Fig.~\ref{fig:motivation}(a), across different operations and batch sizes, a substantial fraction (up to 83.9\%) of RDMA requests become post-failure—i.e., they have already executed at the responder when the failure occurs. Moreover, in a high-throughput setting where 16 clients continuously transfer 64 KB objects in batches of 64 (4 MB per request), uniform retransmission results in a 2.8× longer failover than retransmitting only the pre-failure subset (Fig.~\ref{fig:motivation}(b)). This redundant traffic directly inflates failover latency.

\textbf{2) Duplicate execution and semantic violations.} Blindly resending post-failure requests causes the responder to execute them twice. This jeopardizes applications relying on one-sided semantics.
For example, consider a database client updating a value from A → B using an RDMA write, followed by another client modifying the value from B → C. If the first client experiences a post-failure and blindly retransmits, the stale write B may overwrite the correct value C, violating transactional correctness.

\subsection{Failure-type Aware Retransmission}
To be practical at datacenter scale, a failure-type–aware RDMA failover mechanism is needed to address both the correctness and efficiency shortcomings of blind retransmission. Achieving this goal requires overcoming three key challenges:

\textbf{C1. Correctly recognizing completion without hardware changes. }In RDMA, the canonical signal that an operation has completed is an ACK from the responder. However, commodity NICs do not expose historical ACK state to software. Thus the requester must maintain additional completion metadata in software (or piggybacked on requests) that faithfully reflects the responder’s commit point — without modifying hardware, while keeping the metadata’s runtime and memory overheads minimal and ensuring metadata consistency with actual execution.

\textbf{C2. Reconstructing post-failure return values.} Some post-failure requests (those already executed by the responder) have no ACK visible to the requester, yet their return values are necessary for correctness—for example, a successful CAS must return the value that was present at the instant of the atomic operation. Reconstructing those return values is harder than tracking simple completion: CAS is atomic, so ordinary reads before or after the CAS cannot reveal the value at the CAS execution point. The requester therefore needs a lightweight, externally visible mechanism to retrieve the exact post-failure result of atomic/non-idempotent operations.

\textbf{C3. Fast, resource-efficient failover (time–space tradeoffs).}
Failover requires moving outstanding traffic to a backup path. Existing approaches either pre-allocate RCQPs on standby NICs (sacrificing memory and initialization cost) or create RCQPs on demand (incurring ms-scale stalls). LubeRDMA~\cite{luberdma} reserves a pool of pre-allocated RCQPs across all available links to ensure fast switchover, whereas Mooncake~\cite{mooncake} restricts the number of active RCQPs and creates new RCQPs on demand when a link switch occurs. Both approaches, however, either incur substantial memory overhead or suffer from increased failover latency.

\section{\sysname Design}
\label{sec:design}

\sysname is a runtime RDMA failover framework designed to solve three challenges from Sec.~\ref{sec:motivation}. In this section we will give a architecture overview and detailed explanation of three key mechanisms that together solve these challenges: Compact Completion Logging for failure type identifying (C1), Extended Status for deterministic recovery of post-failures (C2), and DCQP Pools for immediate failover (C3). 

\begin{figure}[!t]
 \centering
 \includegraphics[width=0.49\textwidth]{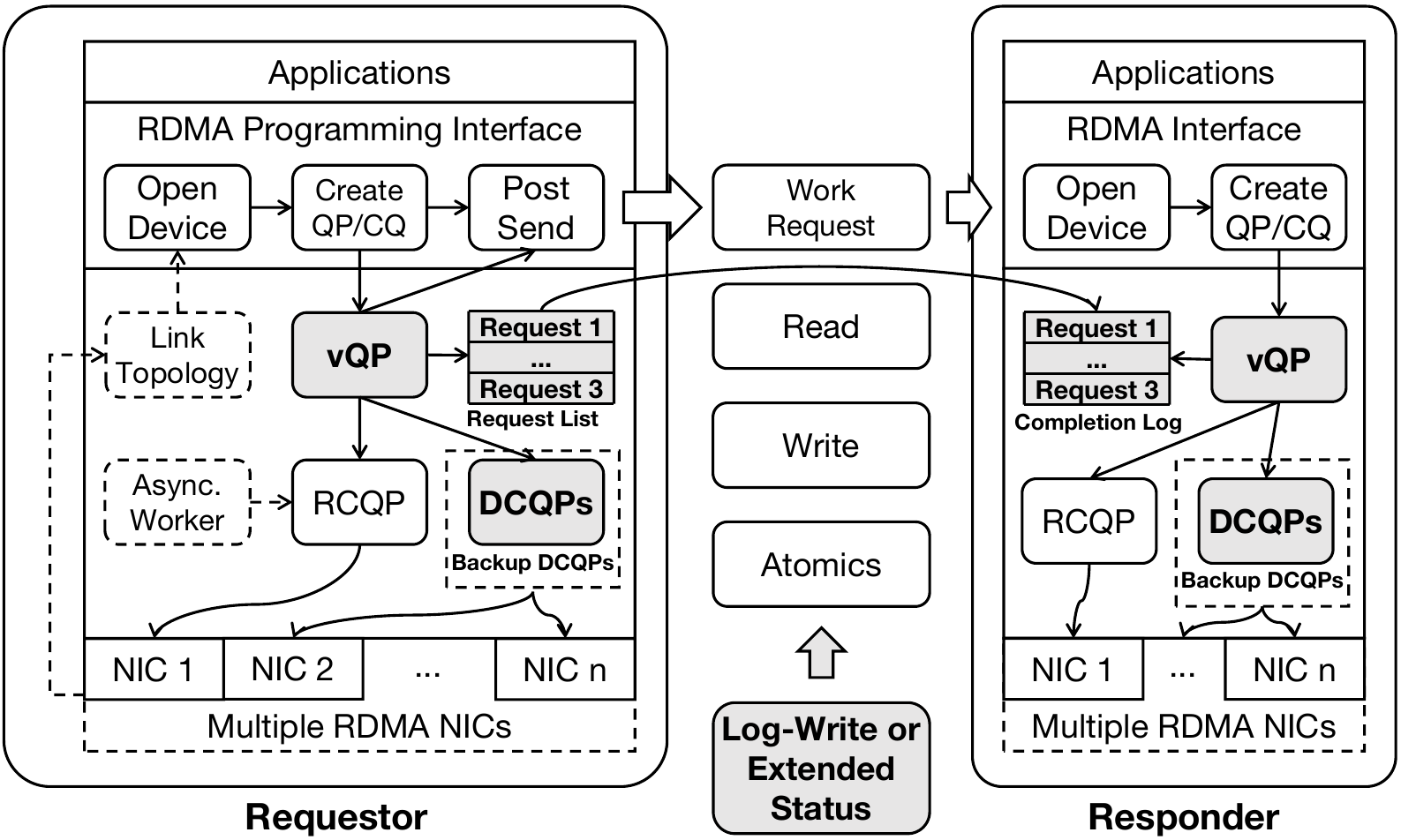}
 \caption{\sysname Overview: Completion logs and extended status provide durable evidence of RDMA failures, and vQPs backed by lightweight DCQPs enable immediate failover.}
\vskip -10pt
 \label{fig:overview}
\vskip -5pt
\end{figure}

\subsection{Architecture}

\sysname is implemented as an RDMA network programming library that sits between applications and the RDMA verbs layer, exposing a standard RDMA API and requiring minimal application changes. Internally (Fig.~\ref{fig:overview}), \sysname comprises three core components:
a) a logical-to-physical connection table that maps each virtual QP (vQP) to one primary RCQP and multiple backup DCQPs across primary and standby links;
b) per-connection request-log and completion-log regions maintained across the requestor and responder; and
c) extended completion records that track per-operation execution status for correct recovery.

\sysname supports common RDMA programming abstractions—including QP management, protection-domain and memory-region management, connection setup, post send, post batch, poll, and others—while transparently integrating its failover mechanisms. It augments RDMA work requests with lightweight completion logging and execution status, dispatches them through available RDMA paths, and automatically triggers failover procedures when detecting failures. 

Algorithms~\ref{alg:v_post_send} and \ref{alg:v_poll_cq} illustrate the simplified workflows of \sysname’s two fundamental APIs and highlight the core design principles of the system. In Algorithm~\ref{alg:v_post_send}, when the application issues a vQP and a set of work requests, \sysname first resolves the currently active physical QP; if the RCQP is undergoing (re)connection, \sysname temporarily falls back to the associated DCQP (Sec.~\ref{sec:failover}). It then augments the application’s work requests with external completion logging and extended execution status for failure-type identification (Sec.~\ref{sec:log}) and return-value recovery (Sec.~\ref{sec:status}), before invoking the raw post-send operation.

After the post-send/poll stage, the post (Alg.~\ref{alg:v_post_send}) and poll (Alg.~\ref{alg:v_poll_cq}) workflows share a common structure: both check the returned status, trigger link switching upon detecting failures (Sec.~\ref{sec:failover}), perform recovery for in-flight requests (Sec.~\ref{sec:log} and Sec.~\ref{sec:status}), and finally update the local request log by inserting or removing entries as needed.

In the following sections, we detail \sysname’s designs for completion logging, execution-status extensions, and DCQP-acceleration fast failover.

\begin{algorithm}[!t]
\caption{Post\_Send}\label{alg:v_post_send}
\begin{algorithmic}[1]

\Procedure{Post\_Send}{vqp, input\_wr}
    \State qp $\gets$ vqp.get\_current\_qp()
    \If{qp.status == CONNECTING}
        \State qp $\gets$ vqp.get\_dcqp()
    \EndIf
    \State wr $\gets$ WR\_Logging(input\_wr)
    \State wr $\gets$ WR\_Extension(wr)
    \State result $\gets$ Raw\_Post\_Send(qp, wr)
    \If{result == error}
        \State \Call{Switch\_VQP}{vqp}
        \State \Call{Recovery}{vqp}
    \EndIf
    \State vqp.append\_log(input\_wr)
\EndProcedure

\end{algorithmic}
\end{algorithm}

\begin{algorithm}[!t]
\caption{Poll\_CQ}\label{alg:v_poll_cq}
\begin{algorithmic}[1]

\Procedure{Poll\_CQ}{vqp}
    \State Raw\_Poll\_CQ(vqp.cq, completion)
    \If{completion.status == error}
        \State \Call{Switch\_VQP}{vqp}
        \State \Call{Recovery}{vqp}
    \EndIf
    \State vqp.remove\_log\_by\_id(completion.id)
\EndProcedure

\end{algorithmic}
\end{algorithm}


\subsection{Log-based Fail Recognition}
\label{sec:log}

\sysname leverages a dual-log design—a request log on the requestor and completion log on the responder—to precisely determine each requests' failure type. Realizing this design requires addressing three key challenges: 1) how to identify and reload request contents under constrained request-log space, 2) how to maintain a correct responder-side completion-log  with minimal performance overhead, and 3) how to correctly handle a batch of RDMA operations when a link failure occurs in the middle of the request sequence.

\begin{figure}[t]
 \centering
 \includegraphics[width=0.4\textwidth]{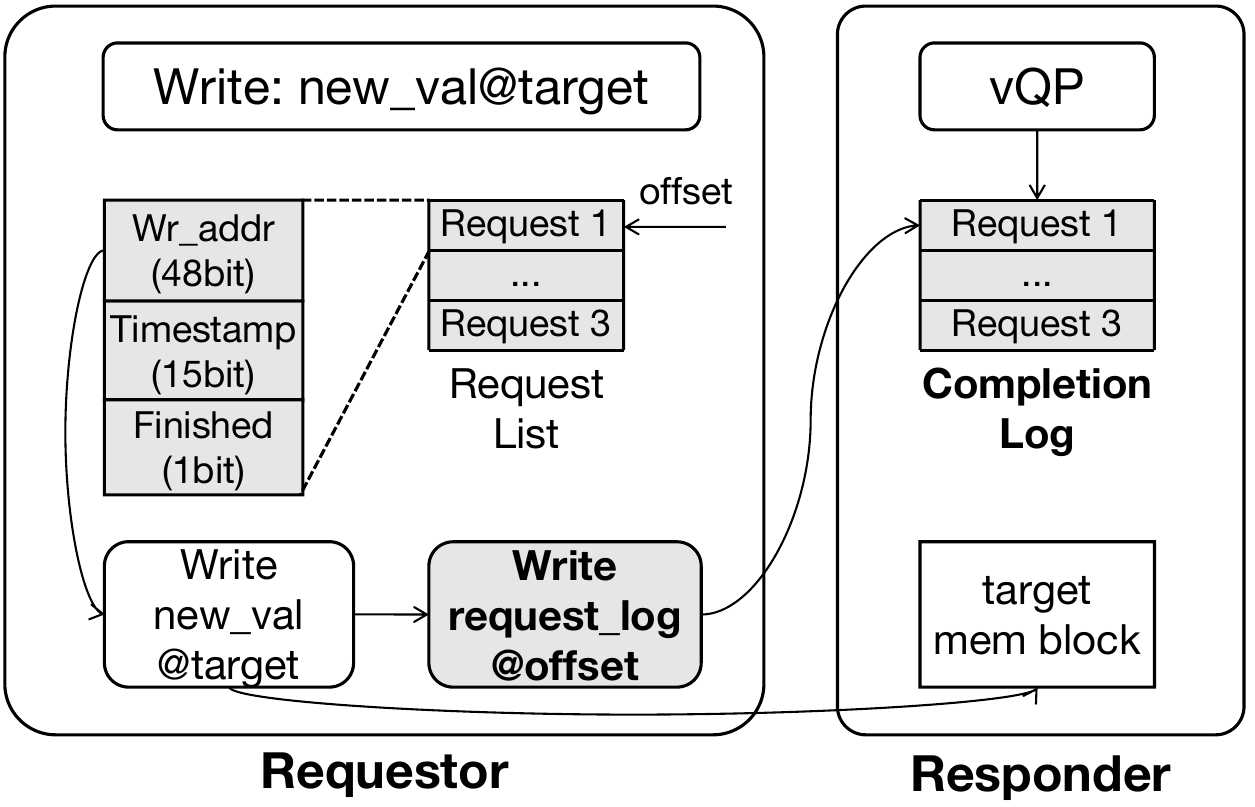}
  \vskip -5pt
 \caption{Completion Log: \sysname records a log at the responder for each request to distinguish pre-failure operations. }\label{fig:write}
\vskip -10pt
\end{figure}

\noindent{\underline{\textbf{1) Unified request identification with request storage.}}} \sysname allocates two per-vQP logs—a request log on the requestor and a completion log on the responder. Both logs share the same 8-byte entry format (Fig.~\ref{fig:write}). 

Before posting a request to the NIC, \sysname first creates a full copy of the request’s metadata (i.e., its struct ibv\_send\_wr) and stores the pointer to this copy in the lower 48 bits of the request-log entry. The remaining bits encode a 15-bit timestamp and a 1-bit finished flag. After appending the log entry, \sysname posts the original request as usual.

Upon receiving a successful completion event, the requestor scans the request log, matches the completed request ID with the corresponding log entry, marks the entry as finished, and frees the copied work request. For applications that do not supply unique request IDs (e.g., always using 0), \sysname uses the 8-byte log entry itself as a unified request identifier, combining its wr\_addr and timestamp.

During recovery, \sysname scans the entire request log and matches each unfinished local entry with its counterpart in the remote completion log. This unified identification relies on comparing the timestamp and the pointer-encoded request metadata. If the timestamps match, the request must have completed before the failure; if they differ, the request was not finished and must be retransmitted. For each unfinished entry, \sysname retrieves the saved ibv\_send\_wr via the stored pointer and resends the corresponding operation.

\noindent{\underline{\textbf{2) Compacted and lightweight completion-log design.}}} \sysname allocates a responder-side completion log for each vQP, which is updated exclusively by the requestor via one-sided RDMA writes.

For every work request issued by the application, \sysname appends a one-sided inline write that updates the corresponding completion-log entry with the same timestamp and request-metadata pointer recorded in the request log. To maintain a single completion event per request, \sysname reuses the original request’s ID for the inline write and transfers the completion-signaling flag from the original operation to the log-write. Thus, only the log-write is configured to generate a completion event.

The correctness of the completion log is guaranteed in two cases.
First, \textbf{for pre-failure requests}, the completion-log write is never executed. Because the inline write is ordered immediately after the actual operation, any failure that interrupts the original request will also prevent its log-write from reaching the responder. As a result, the corresponding completion-log entry remains empty, allowing the recovery process to correctly categorize the request as unexecuted and safely retransmit it.

Second, \textbf{for post-failure requests}, the completion-log write is successfully executed before the failure, ensuring that the request is recorded as completed. A rare corner case exists where the original operation succeeds but the inline log-write fails. We treat this as a low-probability event in practice: the inline write is only 8 bytes and its send window is significantly shorter than the end-to-end execution window of the actual operation. Moreover, once the corner case happened, only one operation will be influenced and it will be treated as a retransmit request, falling back to basic methods.

The overhead of the completion-log writes is negligible. They share the same doorbell with the original send requests, incur no additional round-trips, and can be issued in parallel with application requests, becoming sequential only during NIC execution to keep correctness. When failure happens, the whole completion-log can be fetched by one RDMA read as the size of completion-log is also small (8B per entry).

\noindent{\underline{\textbf{3) Batch operations and asynchronous transport.}}} RDMA applications commonly issue multiple work requests in a single linked list, or dispatch requests asynchronously without waiting for completions—often polling completions in a background thread. \sysname fully supports both batching and asynchronous transports.

For a batch of work requests, \sysname first iterates through the linked list, replicates each work request, and appends a completion-log write after every request. Each request in the batch is inserted into both the request log and the completion log independently, as link failures may occur in the middle of the batch. During recovery, \sysname replays the batch in the order of originally posted, starting from the first pre-failure request, ensuring consistent semantics.

Asynchronous transports are naturally supported because the request log already records all in-flight requests. \sysname exposes completion-queue polling interfaces that allow applications to filter by request ID. And the recovery procedure can be triggered after a poll-completion failure detected.

To further reduce logging overhead, \sysname logs only non-idempotent operations by default (e.g., WRITE, CAS, FAA), as resending these operations incorrectly can violate RDMA semantics. Finally, \sysname provides a pre-allocated work-request pool, allowing applications to directly issue requests using these managed objects without additional memory copies—\sysname safely manages their lifetimes.

\begin{figure}[!t]
 \centering
 \includegraphics[width=0.49\textwidth]{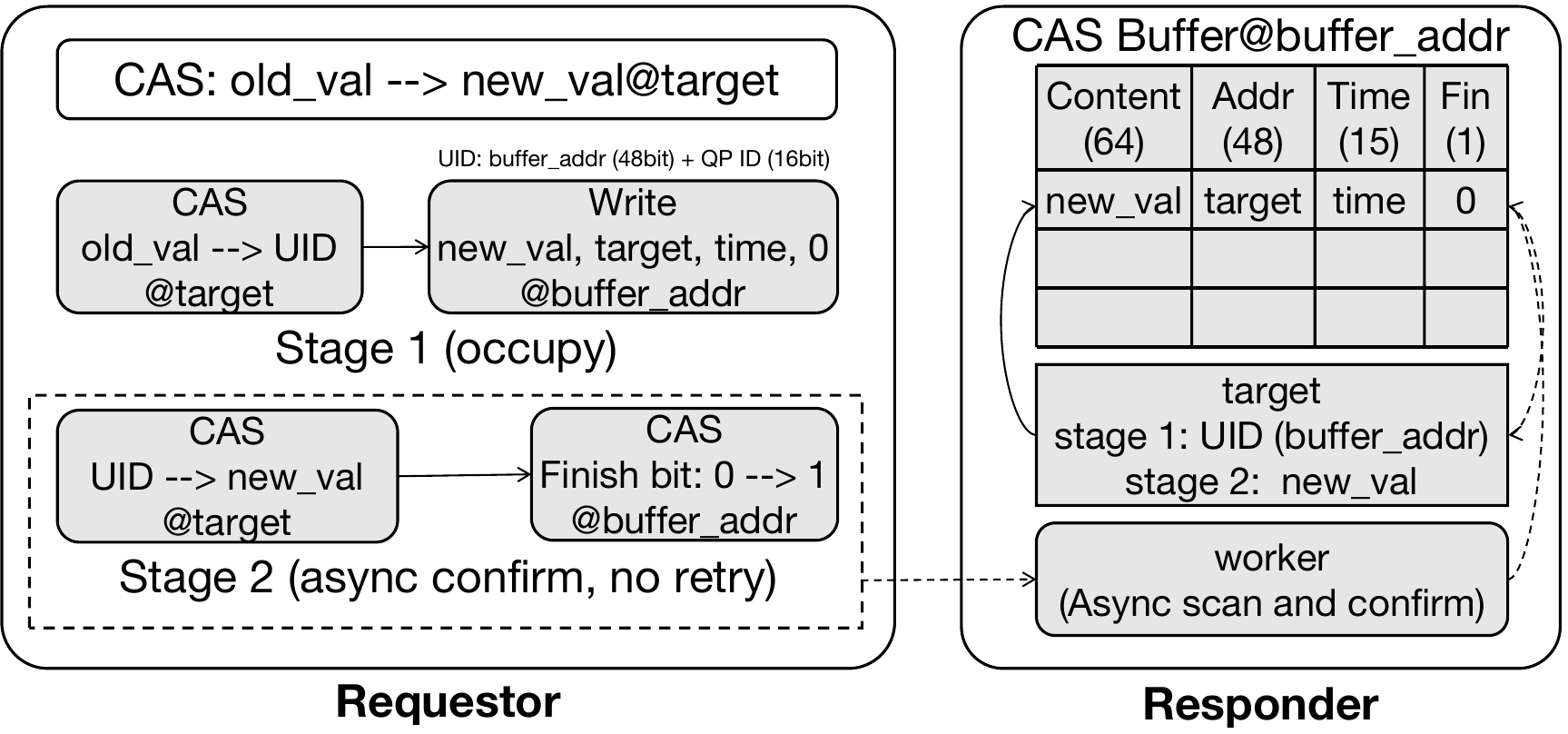}
  \vskip -5pt
 \caption{Extended Status: \sysname implements traceable two-stage atomic operations to replace the original.}\label{fig:atomic}
\vskip -15pt
\end{figure}

\subsection{Post-Failure Recovery via Extended Status}
\label{sec:status}

With the completion-log mechanism, \sysname can identify post-failure requests—operations that have already executed on the responder NIC but whose return values are lost (manifesting as retry errors or timeouts). 
    This enables \sysname to provide return-value recovery, which is particularly important for CAS operations. CAS return values determine whether a lock acquisition or a transaction commit succeeds, and incorrect duplication can break application semantics. To support correct recovery, \sysname extends CAS work requests with additional status information and restructures atomic execution into two stages (Fig.~\ref{fig:atomic}).
    
\noindent{\underline{\textbf{1) Step 1—Occupy. }}} For each CAS operation, the requestor first writes the operation’s payload (the “swap value”), together with the completion-log, into a small per-vQP CAS buffer at the responder. It then constructs a 64-bit UID from the buffer’s address and the requestor’s QP ID (e.g., 48-bit buffer address || 16-bit QP ID). The CAS instruction uses this UID as the swap value instead of the actual value. A successful CAS installs this UID at the target location, making the operation globally unique and unambiguously recoverable: anyone seeing the UID can decode it to locate the buffer and retrieve the actual value and recorded status.
    
\noindent{\underline{\textbf{2) Step 2—Confirm. }}} After a successful CAS, the requestor asynchronously resolves the UID by reading the buffer, marking the operation as finished in the local/remote metadata, and replacing the UID in the target with the actual value via another CAS. To accelerate cleanup, the responder runs a lightweight background worker that scans the CAS buffer and automatically resolves any unflushed completion operations.

\noindent{\underline{\textbf{3) Recovery.}}} Upon failure, \sysname determines the outcome of a CAS by inspecting the target and its metadata: presence of a UID (or a completion record) indicates that the CAS executed and returned success; presence of an unfinished completion record without a UID indicates the CAS executed but returned false; absence of both indicates the operation does not executed and is safe to retransmit. This two-stage pattern (buffer$\rightarrow$UID$\rightarrow$final value) ensures that every successful CAS is provably traceable with metadata compact.

\noindent{\underline{\textbf{4) Correctness.}}} Different from completion-log writes, CAS operations with extended status provide stronger consistency guarantees. \sysname first writes the swap value and the associated completion log, and only then performs the CAS commit point. This ordering ensures that, if the CAS request is not executed or fails, its log entry remains in an unfinished state, prompting the requestor to safely resend the operation. Conversely, if the CAS executes successfully, the corresponding log entry is marked as finished, which reliably prevents any further retransmission. As a result, the extended-status CAS enables \sysname to achieve absolute correctness in distinguishing pre-failure requests (which must be replayed) from post-failure requests (which must not be repeated).

\sysname also exposes an interface that lets applications disable the extended-status mechanism when unnecessary—for example, when applications use globally unique swap-values or when only a single writer can commit the CAS. In these cases, \sysname can simply re-read the target value and compare it to the swap-value to determine the correct return value.

Beyond CAS, \sysname handles other post-failure operations as follows: \textbf{Reads}: safely re-issued, since they are idempotent;  \textbf{Writes}: no action required, as they have already completed; \textbf{Fetch-and-Add}: rewritten into a read+CAS sequence by default, and applications may optionally declare them idempotent for blind retransmission; \textbf{Two-sided Operations}: treated as non-idempotent with no recoverable return value, and applications may mark them idempotent if retransmission is safe.

\subsection{Lightweight Fast Failover}
\label{sec:failover}

\begin{figure}[!t]
 \centering
 \includegraphics[width=0.49\textwidth]{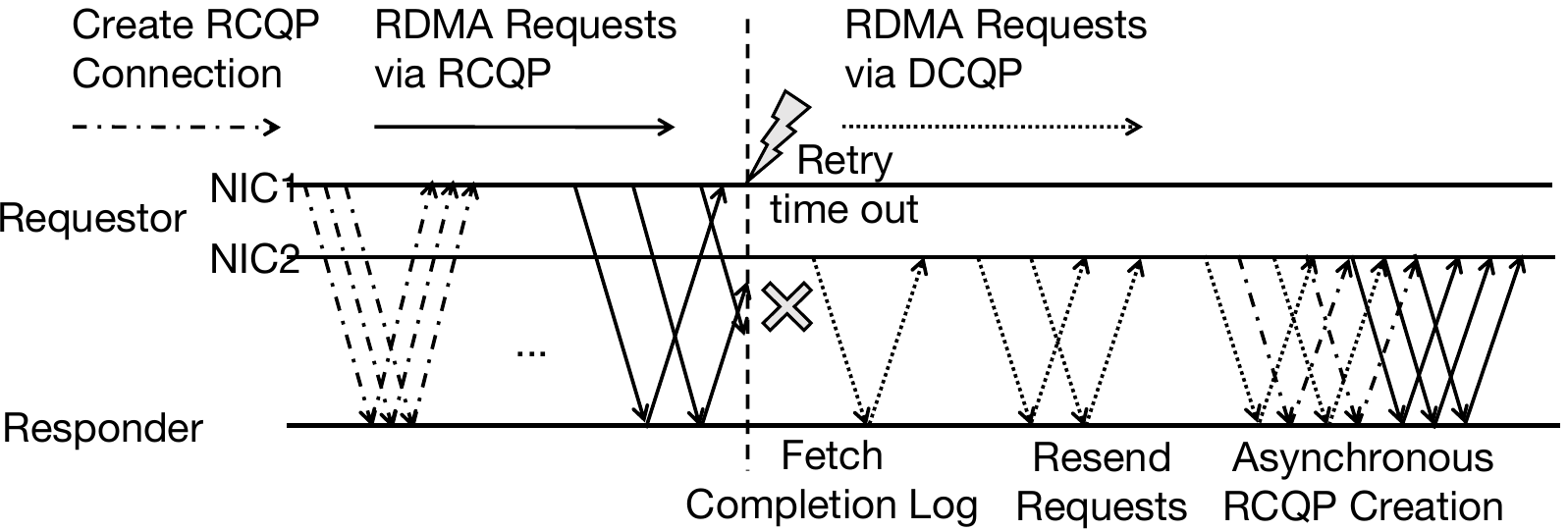}
 \caption{Failover Timeline: \sysname first uses DCQPs as temporary connections to fetch logs and retransmit pre-failure requests, while rebuilding RCQPs in the background.}\label{fig:overview_finemem}
 \vskip -5pt
 \label{fig:path}
\end{figure}

Finally, \sysname assembles a complete failover pipeline using the failure-recognition and recovery mechanisms described above. This section addresses a fundamental performance trade-off in RDMA failover: \textbf{recreating thousands of per‑connection RCQPs on standby links prolongs recovery, whereas fully pre‑caching RCQPs on every link consumes excessive host memory and initialization time}. \sysname resolves this tension by exploiting hardware support for Dynamic Connected QPs (DCQPs). A small, bounded DCQP pool serves as a set of lightweight, shared failover channels, as illustrated in Fig.~\ref{fig:path}. 

\noindent{\underline{\textbf{1) Failover with DCQP Pools.}}} Upon detecting a NIC failure, \sysname atomically remaps the logical→physical QP table so that affected vQPs redirect to available DCQPs on a healthy RDMA link. 
    The current selection policy assigns DCQPs at random—achieving near-uniform sharing under steady load without enforcing strict fairness.  
    Because DCQPs are pre‑allocated and the remapping is a purely local, in‑memory operation that can be paralleled, traffic resumes within milliseconds rather than waiting for new RCQP establishment.
    The DCQP pool size is a tunable operator parameter that balances steady‑state resource usage against transient contention during failover.

DCQP sharing introduces a temporary performance penalty—individual vQP bandwidth may drop and contention may rise—but \sysname confines this bounded degradation to the brief interval required for asynchronous RCQP reconstruction. Once RCQPs are rebuilt, \sysname switches vQP mappings back to dedicated RCQPs, restoring full per-connection throughput. During this swap-back, in-flight requests issued on a DCQP continue to complete on that DCQP and deliver acknowledgments/results to its CQ, while new requests are posted on the rebuilt RCQP and use its own CQ.

\noindent{\underline{\textbf{2) Failover Progress Summary.}}} Building on the mechanisms mentioned above, \sysname’s recovery proceeds in three phases: normal operation, immediate failover (Alg.~\ref{alg:switch_vqp}), and in-flight request recovery (Alg.~\ref{alg:recovery}).

During normal operation, each vQP is bound to its dedicated RCQP. \sysname issues completion-log writes and extended-status metadata only for non-idempotent operations, keeping the common-case overhead minimal.

Upon detecting an RDMA failure—via driver or firmware callbacks, link-state notifications, completion-queue errors, or heartbeat timeouts—\sysname triggers an atomic, batched remapping of all vQPs associated with the failed link to their corresponding DCQPs on a standby link. Because the logical→physical mapping table is updated in one coherent step, applications never observe partially switched or inconsistent states. Traffic resumes immediately over the pre-allocated DCQPs.
Concurrently, an asynchronous repair process reconstructs each vQP’s RCQP on the standby link, including address exchange and QP state transitions, as shown in Alg.~\ref{alg:switch_vqp}. Once reconstruction completes, \sysname atomically redirects each vQP from the shared DCQP to its newly established RCQP, restoring full per-connection throughput.

During immediate failover, \sysname also initiates recovery logic that inspects responder-side completion logs to determine which in-flight requests require retransmission and which require return-value fetching, as illustrated in Alg.~\ref{alg:recovery}. After recovery completes, the system transparently resumes normal request processing without any application-visible disruption or modification.

\begin{algorithm}[!t]
\caption{Switch\_VQP}\label{alg:switch_vqp}
\begin{algorithmic}[1]
\Procedure{Switch\_VQP}{vqp}
    \State link\_id $\gets$ vqp.next\_available\_link()
    \State vqp.create\_qp\_async(link\_id)
    \State vqp.change\_current\_qp(link\_id)
\EndProcedure

\end{algorithmic}
\end{algorithm}

\begin{algorithm}[!t]
\caption{Recovery}\label{alg:recovery}
\begin{algorithmic}[1]

\Procedure{Recovery}{vqp}
    \State Read(vqp, vqp.remote\_log\_addr, remote\_log)
    \For{i = vqp.log\_start \textbf{to} vqp.log\_end}
        \If{remote\_log[i] == local\_log[i]}
            \State CAS\_Recovery(local\_log[i])
        \Else
            \State \Call{Post\_Send}{vqp, *local\_log[i].wr}
        \EndIf
        \State vqp.remove\_log(i)
    \EndFor
\EndProcedure

\end{algorithmic}
\end{algorithm}

\section{Implementation}
\label{sec:impl}

This section describes our \sysname prototype, highlighting integration points and the engineering choices we made to keep the runtime minimally invasive and compatible with RDMA hardware, drivers and applications. 

\noindent{\underline{\textbf{Integration.}}} \sysname is a runtime layer that intercepts connection management and the RDMA send path while preserving the standard verbs API so existing applications require little source code changes. The layer can be deployed as a user‑space library wrapper (we target a shim for libibverbs) or, when tighter integration is desired, as a thin kernel module or driver. \sysname relies only on standard RDMA primitives, libibverbs, and unmodified kernel drivers.


\noindent{\underline{\textbf{Failure Detection.}}} 
\sysname aggregates failure notifications from multiple complementary sources. Firmware- and driver-level events, along with RDMA completion errors, provide the primary detection signals, while a lightweight control-channel heartbeat offers a robust fallback path. The detection policy—heartbeat interval, retry thresholds, and related parameters—is fully configurable. \sysname also exposes interfaces that allow user-defined failure detectors to update link status or explicitly trigger and revoke failover actions. 

\noindent{\underline{\textbf{Link Selection.}}} \sysname reads user-defined configuration files that specify the primary and backup RDMA links, as well as the policy for selecting recovery paths. Upon a primary-link failure, \sysname selects backup links in the order defined by the configuration—similar to prior RDMA transfer engines\cite{mooncake}. When the primary link later recovers, \sysname rebuilds RCQPs on that link and transparently migrates subsequent transmissions back to the primary path.

\noindent{\underline{\textbf{DCQP Management.}}} At startup \sysname pre‑allocates a small DCQP pool on each NIC; the pool size is configurable, the prototype uses a practical default of 1 per NIC, but operators can increase this to tens depending on workload, or setting the DCQP number auto-scaling with a DCQP-RCQP ratio, e.g., 1:8 means adding 1 more DCQP after 8 RCQP created. Before a DCQP can be used to communicate with a remote endpoint, \sysname must resolve and obtain an Address Handle (AH) for that endpoint. Creating an AH is relatively expensive, so \sysname generates AHs lazily the first time an RCQP to that endpoint is established and then caches the AH for subsequent use by DCQPs or newly created RCQPs.


\noindent{\underline{\textbf{Memory Management.}}} \sysname uses a simple but effective memory‑registration policy to support multi‑NIC access. For each application memory region, the runtime registers the region on each active NIC and records the resulting per‑NIC MR and rkey. These per‑NIC rkeys are stored in a small lookup table indexed by (region id, NIC id). When a responder shares access information with a requester (for example, during connection setup), it includes the set of per‑NIC rkeys; the requester keeps this rkey set and selects the appropriate rkey when sending requests via a particular NIC. This approach lets \sysname transparently target different remote NICs without re‑registering memory at failover time.

\section{Evaluation}
\label{sec:evaluation}

We evaluate the performance and resource overhead of \sysname, benchmarking RDMA applications using \sysname against leading state-of-the-art solutions. Our evaluation is structured to address the following key questions:
\begin{itemize}
    \setlength\itemsep{0em}
    \setlength\parsep{-0.4em}
    \item \textbf{Steady-state performance:} How does \sysname operate when no failures occur? What is the runtime cost of maintaining completion logs and extended-status metadata?
    \item \textbf{Recovery efficiency:} How does \sysname perform during RDMA failures? What are the impacts on bandwidth due to retransmission and RCQP reconstruction?
    \item \textbf{Transactional integration:} How does \sysname affect end-to-end performance for representative RDMA transactional workloads (e.g., TPC-C)? 
\end{itemize}

\subsection{Experimental Setup}

Our testbed consists of 4 servers with 16‑core CPUs, 128GB RAM, and dual 25 Gbps RDMA NICs over PCIe 4.0. The 4 servers connect with each other in via two Ethernet switches, thus they have two links to connect to each other. The software stack uses Linux kernel 5.4.0 and latest RDMA NIC driver.

We compare \sysname against three baselines from state-of-the-art work designs and implement these policies in \sysname code framework to keep performance consistency.

\noindent{\underline{\textbf{No-backup.}}} represents standard RDMA usage without any recovery support, and without external mechanisms such as request logs or completion logs.

\noindent{\underline{\textbf{Resend.}}} maintains a local request log, synchronously rebuilds all affected RCQPs on a standby NIC after a failure, and blindly retransmits all in-flight requests.

\noindent{\underline{\textbf{Resend-cache.}}} is similar to Resend in that it uses a local request log, but it pre-establishes and maintains duplicate RCQPs on both the primary and backup links to avoid reconstruction delays. We only show this baseline at recovery-related evaluations, other cases it is the same with Resend.

Our evaluation consists of two parts: multi-to-one inbound microbenchmarks and multi-node cluster experiments for end-to-end RDMA transaction performance. In the microbenchmarks, the requestor side send up to 16 clients issuing one-sided RDMA operations of varying request sizes to a single server node. For end-to-end evaluation, we build an 4-node RDMA transaction system atop Motor’s three-replica RDMA transaction framework~\cite{motor} and use TPC-C as the workload.



\subsection{Steady-state Performance}
\label{sec:eval_resource}

We characterize the steady-state performance of RDMA transport adapting \sysname and other baseline methods, to show the normal influence of \sysname' design with no RDMA failure happens, which is a more common cases in datacenters. 

We evaluate \sysname along several dimensions:
(i) how its performance scales with different transport payload sizes, and
(ii) how it behaves under varying degrees of concurrency.
We measure both latency and bandwidth under two request patterns—synchronous and batched—while varying the number of concurrent clients.
In the synchronous mode, each client issues a single request at a time and waits for its completion before issuing the next. In the batched mode, multiple operations are aggregated into a single work-request list, generating one completion entry per batch; this reduces posting overhead while preserving ordering semantics. We use a batch size of 64 packets, a common configuration in AI data-transfer engines \cite{mooncake} and database systems \cite{fusee}.
Unless otherwise stated, all experiments use inbound workloads where multiple clients issue one-sided RDMA operations to a single server.

\begin{figure}[!t]
 \centering
 \includegraphics[width=0.48\textwidth]{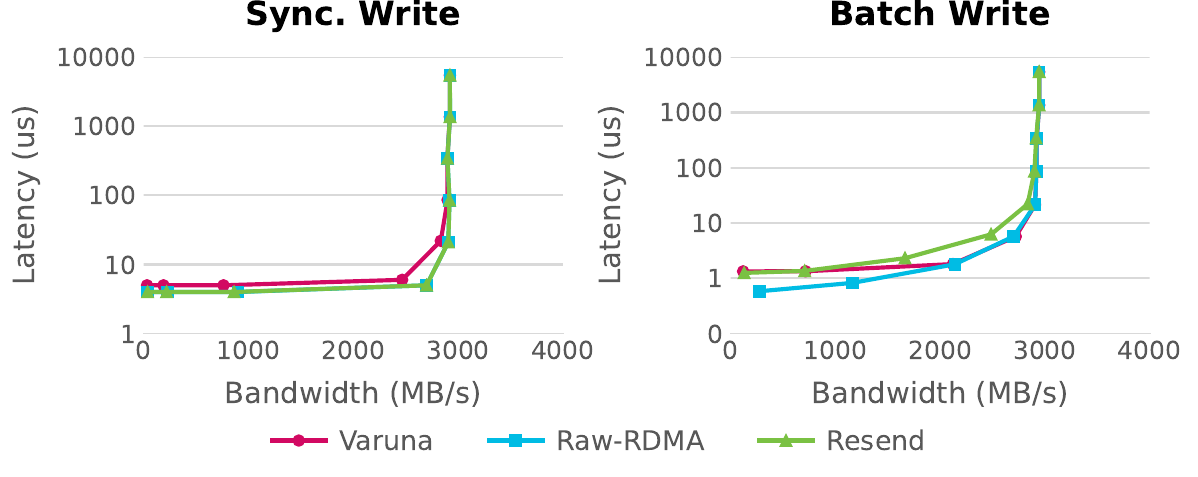}
 \caption{Synchronous and batched writes with payload sizes ranging from 16B to 1MB, using 16 client threads and 64 packages per batch.}
 \label{fig:size}
   \vskip -5pt
\end{figure}

\noindent{\underline{\textbf{Performance under Ranging Payloads.}}} First, we evaluate \sysname using one-sided RDMA writes with payload sizes ranging from 16B to 1MB (Fig.~\ref{fig:size}). All three configurations reach approximately 25Gbps bandwidth at a 4KB payload size, with comparable latencies. Beyond 4 KB, the average latency increases roughly linearly with payload size. Batched writes achieve lower latency than synchronous writes due to reduced posting overhead, and the two modes converge in both latency and bandwidth once payloads exceed 4 KB.
\sysname maintains similar external latency across payload sizes, adding only 1 µs compared to the resend and no-backup baselines—primarily due to log writes. This overhead is largely hidden under batched writes or larger payload sizes. \sysname also sustains the same peak bandwidth as the baselines, even though its internal log-write bandwidth is not counted in the reported throughput.
Overall, for payloads larger than 4 KB—which are common in RDMA deployments, especially when batching is enabled—\sysname incurs at most a 4.7\% external latency overhead and a 2.5\% external bandwidth overhead.



\begin{figure}[!t]
 \centering
 \includegraphics[width=0.48\textwidth]{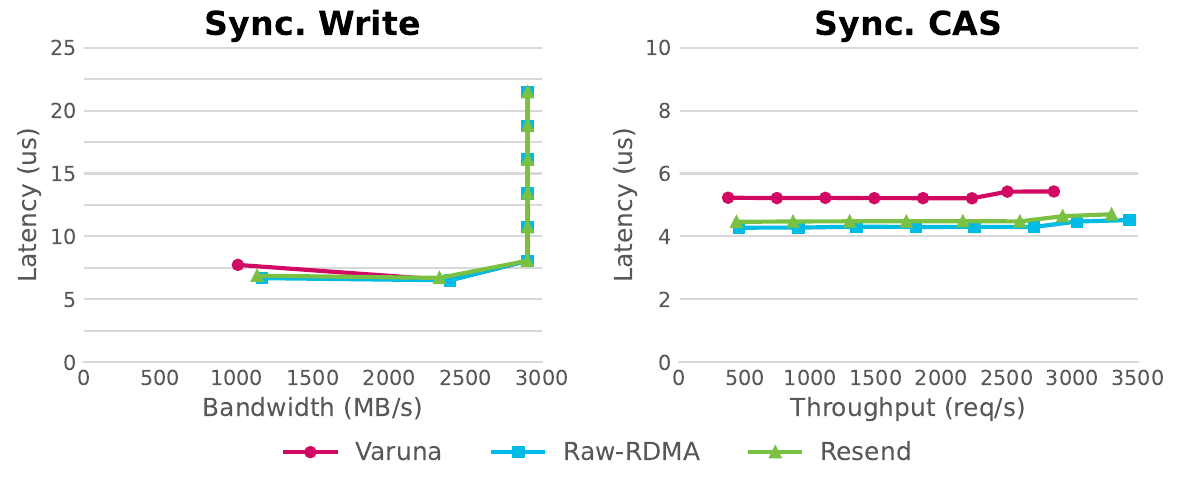}
 \caption{Performance of synchronous RDMA one-sided Write and CAS operations with 1–16 client threads.}
 \label{fig:sync}
   \vskip -5pt
\end{figure}

\noindent{\underline{\textbf{Performance under Ranging Concurrency.}}} Next, we evaluate \sysname using 4 KB writes and 8 B CAS operations under varying concurrency levels, scaling up to 16 clients. For the synchronous pattern, we measure \sysname across two types of one-sided operations—write and CAS—as shown in Fig.~\ref{fig:sync}. We omit RDMA read results because \sysname behaves similarly to the baselines for reads.
\sysname introduces negligible latency and bandwidth overhead for 4 KB writes across all thread counts. In contrast, synchronized CAS operations exhibit relatively higher overhead, as CAS has a very small payload and is executed serially at the responder. However, in real transactional workloads, CAS is typically issued together with one-sided reads in batched form. Accordingly, we present batched-operation evaluations next, where this overhead is largely amortized.

\begin{figure}[!t]
 \centering
 \includegraphics[width=0.48\textwidth]{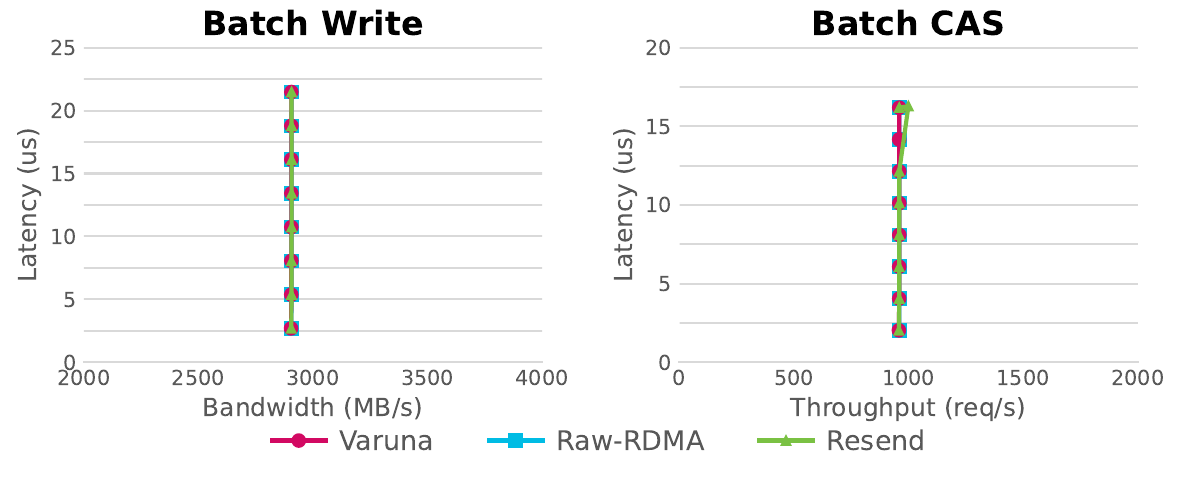}
 \caption{Performance of Write and CAS-Read batches (1:3 ratio) with 1–16 client threads and 64 packages per batch.}
 \label{fig:batch}
   \vskip -5pt
\end{figure}

For the batched pattern, we evaluate \sysname using two types of RDMA batch requests: (i) batches of 4 KB writes, and (ii) batches combining an 8 B CAS followed by three read operations (reflecting typical RDMA transactional locking behavior \cite{motor}), as shown in Fig.~\ref{fig:batch}. Under batching, \sysname achieves latency and bandwidth nearly identical to both the No-backup and Resend baselines. The batching optimization in RDMA effectively amortizes the overhead of \sysname’s log writes, making the added cost almost invisible.
Batch execution is prevalent in RDMA-based systems—for example, AI data-transfer engines commonly use batch sizes of 128 with 64 KB payloads \cite{mooncake}, and transactional workloads frequently issue batches combining one CAS with several reads \cite{motor}. These characteristics further reinforce that \sysname's overhead remains negligible in realistic deployment scenarios.





\noindent{\underline{\textbf{Overhead Drill-down.}}} Breaking down the execution path of a one-sided RDMA write, the pipeline consists of four stages: copying the work request, enqueueing it onto the send queue, transmitting the request together with the in-line log entry, and finally polling for completion. As shown in Fig.~\ref{fig:size}–\ref{fig:batch}, the memory-copy and queue-management stages contribute negligibly compared to the transmission stage. This is reflected in the nearly identical performance of the Resend and No-backup baselines, where Resend performs only local log recording without remote log writes.
The dominant source of additional latency in \sysname comes from its in-line log write: the NIC must complete the log write before issuing the corresponding ACK, extending end-to-end request latency by approximately 1 µs.

\noindent{\underline{\textbf{Memory overheads.}}} \sysname incurs a low steady-state memory overhead through its memory-efficient use of DCQPs, comparable to the no-cache baselines (No-backup and Resend). In contrast, Resend-Cache roughly doubles memory usage due to RCQP duplication. For example, with 4,096 QPs, Resend-Cache consumes nearly twice the memory of \sysname (3,000 MB vs. 1,500 MB).
On the other hand, \sysname’s request and completion logs add only modest overhead—approximately 1 KB per QP—amounting to about 4 MB out of the 1,500 MB total at 4,096 QPs. 




\subsection{Recovery Efficiency}
\label{sec:eval_recovery}

We evaluate the recovery efficiency of \sysname and the baseline mechanisms by injecting random RDMA failures under the batch communication pattern. In this section, we focus on RDMA Write batch and CAS-Read batch to measure recovery time and bandwidth variance—two metrics that are critical for both AI training and database workloads. We additionally evaluate RDMA CAS operations to verify correctness under failure, ensuring that \sysname’ completion log and external state tracking correctly prevent duplicate execution.

\begin{figure}[!t]
  \centering
  \includegraphics[width=0.48\textwidth]{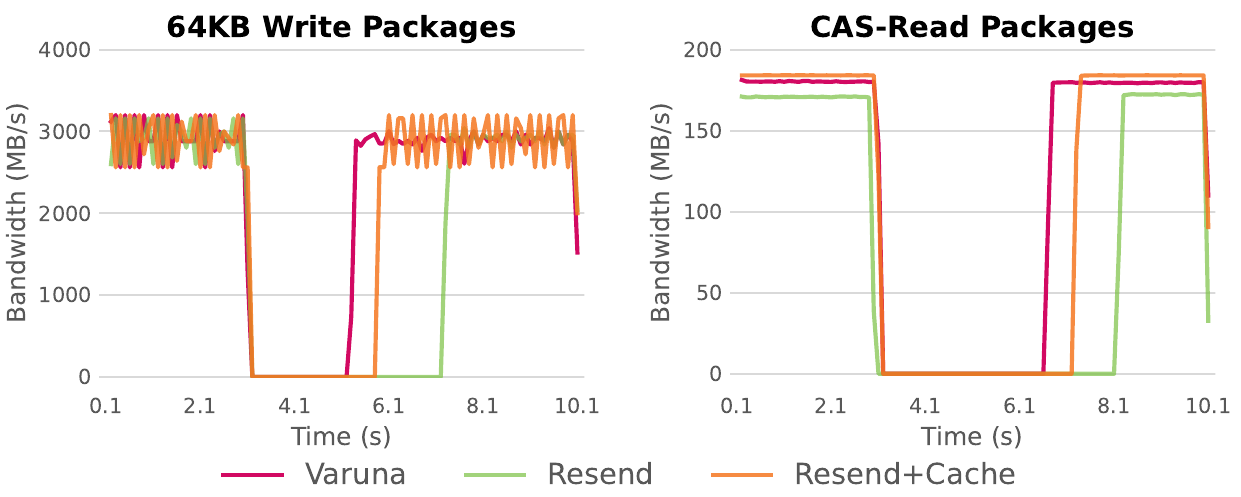}
  \caption{Requestor-side bandwidth during RDMA failover with 16 client threads and 64 packages per batch.}
  \vskip -5pt
  \label{fig:recovery_bandwidth}
  \vskip -5pt
\end{figure}

\noindent{\underline{\textbf{Recovery Bandwidth Consuming.}}} First, we evaluate the recovery bandwidth consumption of resend-based approaches, using 4KB/64 KB payloads and a batch size of 64—an AI-style transmission pattern similar to the Mooncake engine \cite{mooncake}. We compare two retransmission baselines (Resend and Resend-cache) against \sysname. As shown in Fig.~\ref{fig:recovery_bandwidth}, \sysname reduces recovery time by 52.2\% / 64.5\% compared to resend-based mechanisms, which must retransmit all in-flight packets over the backup path, regardless of whether RCQPs are pre-cached.

When accounting for total data transferred during failover, \sysname sends only 25.4\% of the data required by Resend and Resend-cache at 64KB packages, since it retransmits only pre-failure requests. Under batched and large-payload workloads, failed requests often span the middle of a batch list, meaning that only part of the batch was actually transmitted before the failure. \sysname avoids retransmitting the already-completed portion, further reducing recovery bandwidth.



\begin{figure}[!t]
  \centering
  \includegraphics[width=0.48\textwidth]{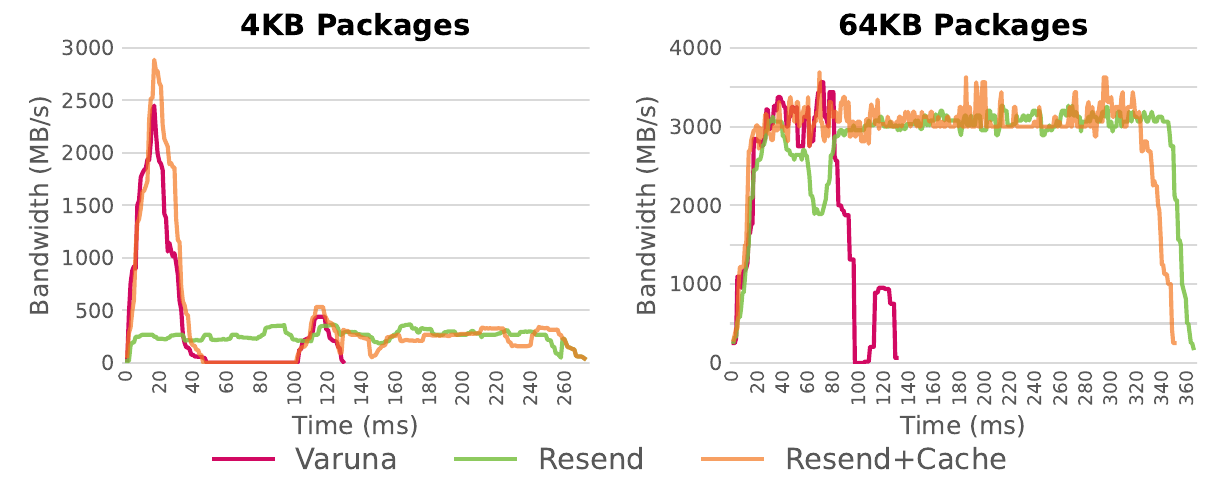}
  \caption{Retransmission bandwidth during RDMA failover with 16 client threads and 64 write packages per batch.}
  \vskip -5pt
  \label{fig:recovery_latency}
  \vskip -5pt
\end{figure}

\noindent{\underline{\textbf{Bandwidth Variance during Recovery.}}} Figure~\ref{fig:recovery_bandwidth} shows a representative throughput time series during failover for both Write and CAS-Read batches (64 packages per batch). Before the injected fault, all systems operate at the same baseline throughput. At the moment of failure, the Resend baseline drops near zero while RCQPs are rebuilt, whereas Resend-Cache maintains throughput close to baseline. \sysname immediately remaps vQPs to DCQPs and reconciles state in the background, sustaining near-baseline throughput throughout the transition and fully returning to baseline once RCQPs are restored. \sysname slightly outperforms Resend-Cache due to reduced retransmission overhead.


\noindent{\underline{\textbf{Correctness.}}} To evaluate correctness under failure, we construct batches of CAS operations encoded as a single work-request list and issue them to the responder while injecting RDMA link failures at random times. Since failures may occur before, during, or after the transmission of the list, each failure naturally generates a mix of pre-failure and post-failure CAS requests. Under this setup, \sysname achieves 100\% correctness. Its two-step CAS protocol, together with extended completion-status tracking, guarantees that all successful CAS operations are reliably recorded and recoverable, without risking duplicate execution.


\subsection{Transaction Integration}

\begin{figure}[!t]
  \centering
  \includegraphics[width=0.48\textwidth]{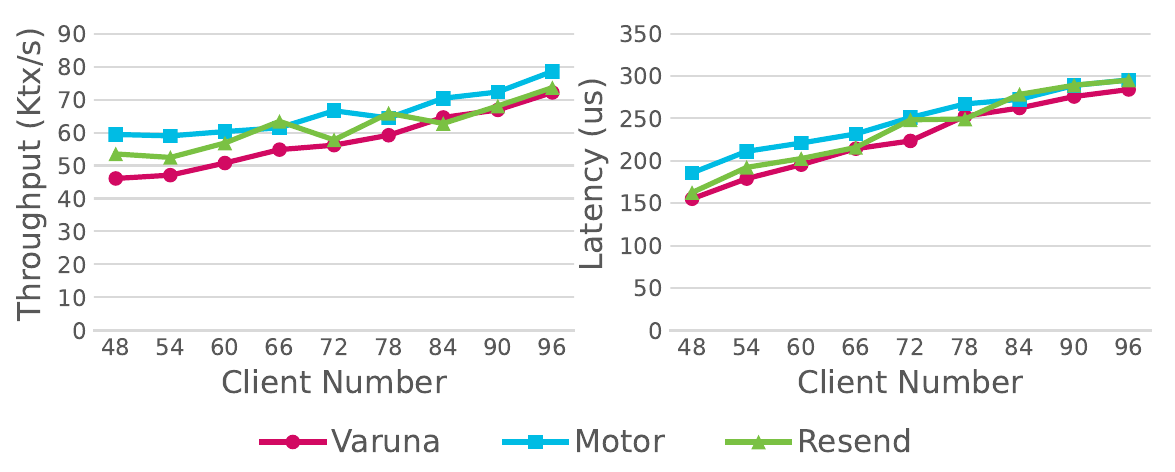}
  \caption{Average latency and throughput of RDMA Transaction Motor \cite{motor} running TPC-C.} 
  \vskip -5pt
  \label{fig:transaction-overehad}
  \vskip -5pt
\end{figure}

\begin{figure}[!t]
  \centering
  \includegraphics[width=0.48\textwidth]{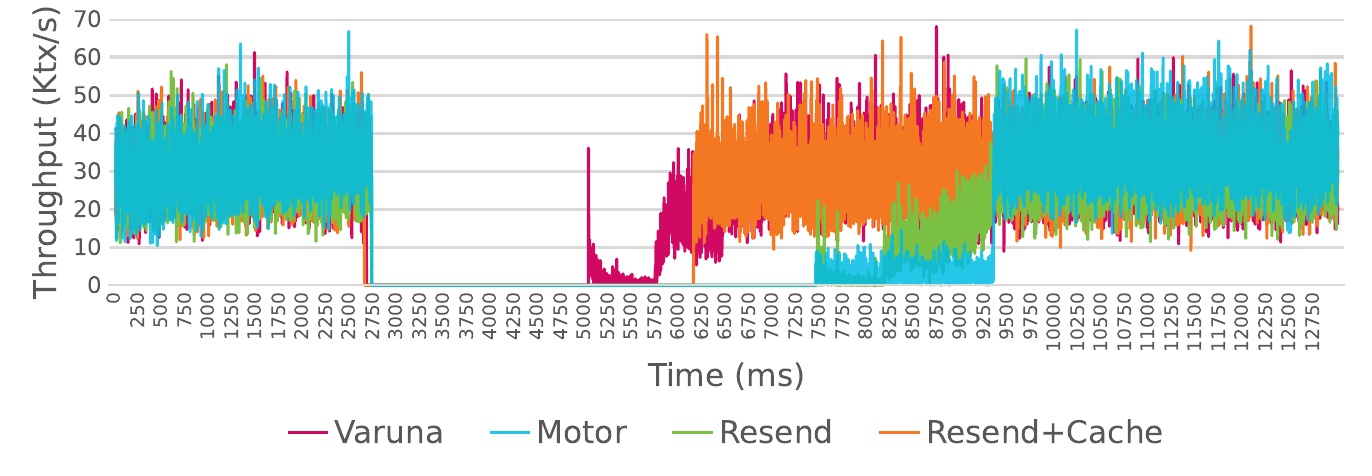}
  \caption{Throughput of Motor under network failure.}
  \vskip -5pt
  \label{fig:transaction-recovery}
  \vskip -5pt
\end{figure}

Finally, we integrate \sysname and other baselines into Motor, an RDMA-based distributed transaction with three memory node replicas. This integration modifies the calling of network interfaces in Motor, with less than 100 LoC modification.

\noindent{\underline{\textbf{Transaction execution performance.}}} As shown in Fig.~\ref{fig:transaction-overehad}, we evaluate \sysname against other baselines using the TPC-C workload. \sysname achieves nearly identical throughput compared to the baselines, indicating that its failover-related mechanisms introduce 1.7\%-13.9\% bandwidth and 0.6\%-10\% average latency overhead in real application scenarios. 

The low overhead primarily stems from the fact that \sysname does not introduce log-writing operations for RDMA reads, which constitute a large fraction of network requests in Motor’s transaction processing. For RDMA writes and CAS operations, the additional log-writing overhead is minimal with batch optimization in Motor. Moreover, the overhead of CAS can be further reduced by Motor’s unified CAS-value modification in its lock design, which naturally amortizes the extra operations required by \sysname.

\noindent{\underline{\textbf{Transaction failover progress.}}} We inject external failures using the same methodology as in prior experiments (Fig.~\ref{fig:transaction-recovery}). \sysname achieves rapid throughput recovery, comparable to both Resend and Resend-cache. By contrast, the Resend baseline suffers additional delay due to RCQP reconstruction. For comparison, we also evaluate Motor’s recovery behavior under a network switch failure—where no packet retransmission occurs and recovery is handled solely at the application layer. We emulate Motor’s external detection delay by inserting a zero-bandwidth interval equivalent to Resend’s failure-detection and RCQP-rebuild time. Motor experiences an even longer recovery period because it relies entirely on application-level mechanisms and must additionally wait for external failure detection before resuming execution.

\noindent{\underline{\textbf{Transaction recovery correctness.}}} Finally, in transaction failover evaluations, \sysname achieves a 100\% correct resubmission success rate with no consistency errors, stalls, or crashes, matching the reliability of application-level recovery. 



\section{Discussion}
\label{sec:limitation}

While \sysname improves recovery performance and deterministic correctness, our design has several limitations:



\textbf{1) Overhead of completion logs and atomic UIDs.} Completion logging and per‑operation UID tracking introduce additional metadata writes. In our measurements these costs are microsecond‑scale per logged request, but workloads dominated by non‑idempotent or atomic operations could incur higher overhead. Adaptive batching, asynchronization, or selective logging policies can reduce this cost in practice. Looking ahead, offloading log persistence to a programmable NIC could further eliminate these costs from the host datapath.

\textbf{2) Background RCQP reconstruction.} DCQPs enable immediate failover, yet rebuilding full RCQPs in the background consumes CPU and driver resources. In environments with very frequent NIC churn, a non‑trivial fraction of system resources may be devoted to RCQP reconstruction; admission or backoff policies may be required to bound reconstruction effort under pathological churn.

\textbf{3) Multi-path selection and load balancing.} \sysname adopts a simple backup-link selection strategy: it parses a user-provided configuration file to determine the primary and standby links. In practice, deployments may offer multiple candidate paths—for example, a CXL-based NIC pool as discussed in Oasis\cite{oasis}—allowing more sophisticated selection or dynamic load balancing. Extending \sysname with adaptive multi-path scheduling and real-time load balancing is an important direction for future work.



\section{Related Work}
\label{sec:relatedwork}

\sysname intersects multiple areas: RDMA recovery and reliability, multipath and redundant RDMA systems, and RDMA‑based database recovery.

\textbf{Redundant RDMA systems.} Besides previous multi-path\cite{hpn7, mprdma} and multi-NIC\cite{fuselink, sirius} researches in RDMA, recent work like Oasis\cite{oasis} designs a multi-NIC pool based on CXL, leveraging idle NICs in existing clusters as backup devices. However, Oasis targets general-purpose network traffic and cannot directly support RDMA semantics. In contrast, \sysname combines a small pool of dynamic connection QPs with per-operation completion logging, enabling immediate continuity and deterministic recovery for RDMA workloads.

\textbf{RDMA recovery and reliability.} At the link layer, many mechanisms provide reliability\cite{reliability, solar}, but they cannot handle seamless failover in scenarios involving multiple NICs, multiple paths, or multiple cards. Libraries such as LubeRDMA\cite{luberdma} and Mooncake\cite{mooncake} provide recovery to support most cases of multiple links at the network programming layer; however, as noted earlier, their retransmission is not strictly correct and may lead to duplicate execution for non-idempotent operations. \sysname differs by providing deterministic classification of in-flight requests into pre-failure and post-failure, enabling correct retransmission and µs-level failover across multiple NICs and paths.

\textbf{RDMA-based application-level recovery. }Systems such as Motor \cite{motor} and FUSEE \cite{fusee} implement application-layer error recovery for RDMA-based disaggregated databases. In parallel, designs based on multi-node memory replicas or erasure coding \cite{carbink, hydra} leverage RDMA to ensure fault tolerance. While effective, these approaches typically incur recovery delays on the order of hundreds of milliseconds and cannot avoid the bandwidth collapse that arises when memory nodes become temporarily unreachable.
\sysname complements these systems by enabling fine-grained, microsecond-scale failover, eliminating redundant retransmissions and sustaining high throughput during link outages.

\section{Conclusion}
\label{sec:conclusion}

We present \sysname, a failure-type aware RDMA recovery framework with correct retransmission and microsecond-level failover. By introducing completion logs and extended status, \sysname ensures correctness for non-idempotent operations, while provides immediate failover with minimal DCQP pre-caching. Our evaluation show that \sysname reduces recovery latency to milliseconds, sustains high bandwidth during failover, and guarantees transactional correctness.
\newpage
\balance
\bibliographystyle{plain}
\bibliography{ref.bib}



\end{document}